\documentclass[aps,prl,showpacs,twocolumn,superscriptaddress,floatfix]{revtex4-2}
\usepackage[utf8]{inputenc}
\usepackage{verbatim}
\usepackage{amsmath}
\usepackage{mathtools}
\usepackage{gensymb}
\usepackage{amssymb}
\usepackage[pdftex]{graphicx}
\newcommand{\figurescale}{1}
\usepackage{soul}
\usepackage{xcolor}
\usepackage[normalem]{ulem}

\usepackage[pdfencoding=auto]{hyperref}
\usepackage{cleveref}
\hypersetup{colorlinks,
	linkcolor={blue!75!black!80!yellow},
	citecolor={blue!75!black!80!yellow},
	urlcolor={blue!75!black!80!yellow}
}

\usepackage{soul}

\begin{document}

\title{Direct Visualization of Defect-Controlled Diffusion in van der Waals Gaps}

\affiliation{Division of Physical Sciences, College of Letters and Science, University of California, Los Angeles, California 90095, USA}
\affiliation{Department of Materials Science and Engineering, Massachusetts Institute of Technology, Cambridge, Massachusetts 02139, USA}
\affiliation{Beijing National Laboratory for Condensed Matter Physics and Institute of Physics, Chinese Academy of Sciences, Beijing 100190, China}
\affiliation{Mark Kac Center for Complex Systems Research, Jagiellonian University, Kraków, Poland}
\affiliation{Research Center for Electronic and Optical Materials, National Institute for Materials Science, 1-1 Namiki, Tsukuba 305-0044, Japan}
\affiliation{Research Center for Materials Nanoarchitectonics, National Institute for Materials Science,  1-1 Namiki, Tsukuba 305-0044, Japan}

\author{Joachim Dahl Thomsen}
\email{j.thomsen@fz-juelich.de}
\affiliation{Division of Physical Sciences, College of Letters and Science, University of California, Los Angeles, California 90095, USA}
\affiliation{Department of Materials Science and Engineering, Massachusetts Institute of Technology, Cambridge, Massachusetts 02139, USA}

\author{Yaxian Wang}
\affiliation{Beijing National Laboratory for Condensed Matter Physics and Institute of Physics, Chinese Academy of Sciences, Beijing 100190, China}

\author{Henrik Flyvbjerg}
\affiliation{Mark Kac Center for Complex Systems Research, Jagiellonian University, Kraków, Poland}

\author{Eugene Park}
\affiliation{Department of Materials Science and Engineering, Massachusetts Institute of Technology, Cambridge, Massachusetts 02139, USA}

\author{Kenji Watanabe}
\affiliation{Research Center for Electronic and Optical Materials, National Institute for Materials Science, 1-1 Namiki, Tsukuba 305-0044, Japan}

\author{Takashi Taniguchi}
\affiliation{Research Center for Materials Nanoarchitectonics, National Institute for Materials Science,  1-1 Namiki, Tsukuba 305-0044, Japan}

\author{Prineha Narang}\email{prineha@ucla.edu}
\affiliation{Division of Physical Sciences, College of Letters and Science, University of California, Los Angeles, California 90095, USA}

\author{Frances M.~Ross}\email{fmross@mit.edu}
\affiliation{Department of Materials Science and Engineering, Massachusetts Institute of Technology, Cambridge, Massachusetts 02139, USA}

\date{\today}

\begin{abstract}
Diffusion processes govern fundamental phenomena such as phase transformations, doping, and intercalation in van der Waals (vdW) bonded materials. Here, we quantify the diffusion dynamics of W atoms by visualizing the motion of individual atoms at three different vdW interfaces: BN/vacuum, BN/BN, and BN/WSe$_2$, by recording scanning transmission electron microscopy movies. Supported by density functional theory calculations, we infer that in all cases diffusion is governed by intermittent trapping at electron beam-generated defect sites. This leads to diffusion properties that depend strongly on the number of defects. These results suggest that diffusion and intercalation processes in vdW materials are highly tunable and sensitive to crystal quality. The demonstration of imaging, with high spatial and temporal resolution, of layers and individual atoms inside vdW heterostructures offers possibilities for direct visualization of diffusion and atomic interactions, as well as for experiments exploring atomic structures, their in-situ modification, and electrical property measurements of active devices combined with atomic resolution imaging.
\end{abstract}

\maketitle

\section{Introduction}
Diffusion of atomic species in and on two-dimensional (2D) materials governs numerous phenomena such as phase transformations \cite{lin2014atomic}, crystal growth \cite{zhang2018diffusion}, and doping \cite{
spear1975substitutional, 
wang2018atomic}. In particular, the diffusion of foreign atoms into van der Waals (vdW) bonded materials to form intercalated structures has great potential for tuning physical properties (electronic, optoelectronic, magnetic), and for uses in energy storage and catalysis \cite{wu2023electrostatic, rajapakse2021intercalation}. 

Of particular interest, yet not widely studied, is diffusion within assembled vdW heterostructures. Such structures, consisting of stacked multilayer blocks of 2D crystals, form the basis for fundamental studies of the electrical properties and physics of 2D materials and devices. Most importantly, 2D materials encapsulated by hexagonal boron nitride (BN) are the de facto standard for creating devices with high-quality electrical properties \cite{rhodes2019disorder}; stacked heterostructures show promise for novel electronic, optoelectronic \cite{liu2016van, novoselov20162d}, magnetic devices \cite{burch2018magnetism, gibertini2019magnetic}, and for exploring phenomena such as strongly correlated electron physics \cite{kennes2021moire}. A microscopic understanding of diffusion pathways and dynamics in vdW heterostructures, including the role of defects, is important in providing opportunities for tailoring the properties of these structures. Measurements of diffusion in such heterostructures also offer advantages compared to studies of atomic motion on surfaces, since surface contamination issues are avoided and the diffusion environment can be well-controlled by trapping diffusing atoms between particular layers within the structure.
For such studies, TEM is a powerful tool with possibilities for real-time, atomic resolution imaging of diffusion processes, and has been used to image and quantify the diffusion of adsorbed or substitutional atoms on the surface of different 2D materials \cite{yang2019direct, gan2008one, hong2017direct, li2017atomic}, diffusion of dopants in a bulk semiconducting crystal \cite{ishikawa2014direct}, and the motion of noble gas clusters ion-implanted into bilayer graphene \cite{langle2024two}. 

\begin{figure*}[]
	\scalebox{\figurescale}{\includegraphics[width=1\linewidth]{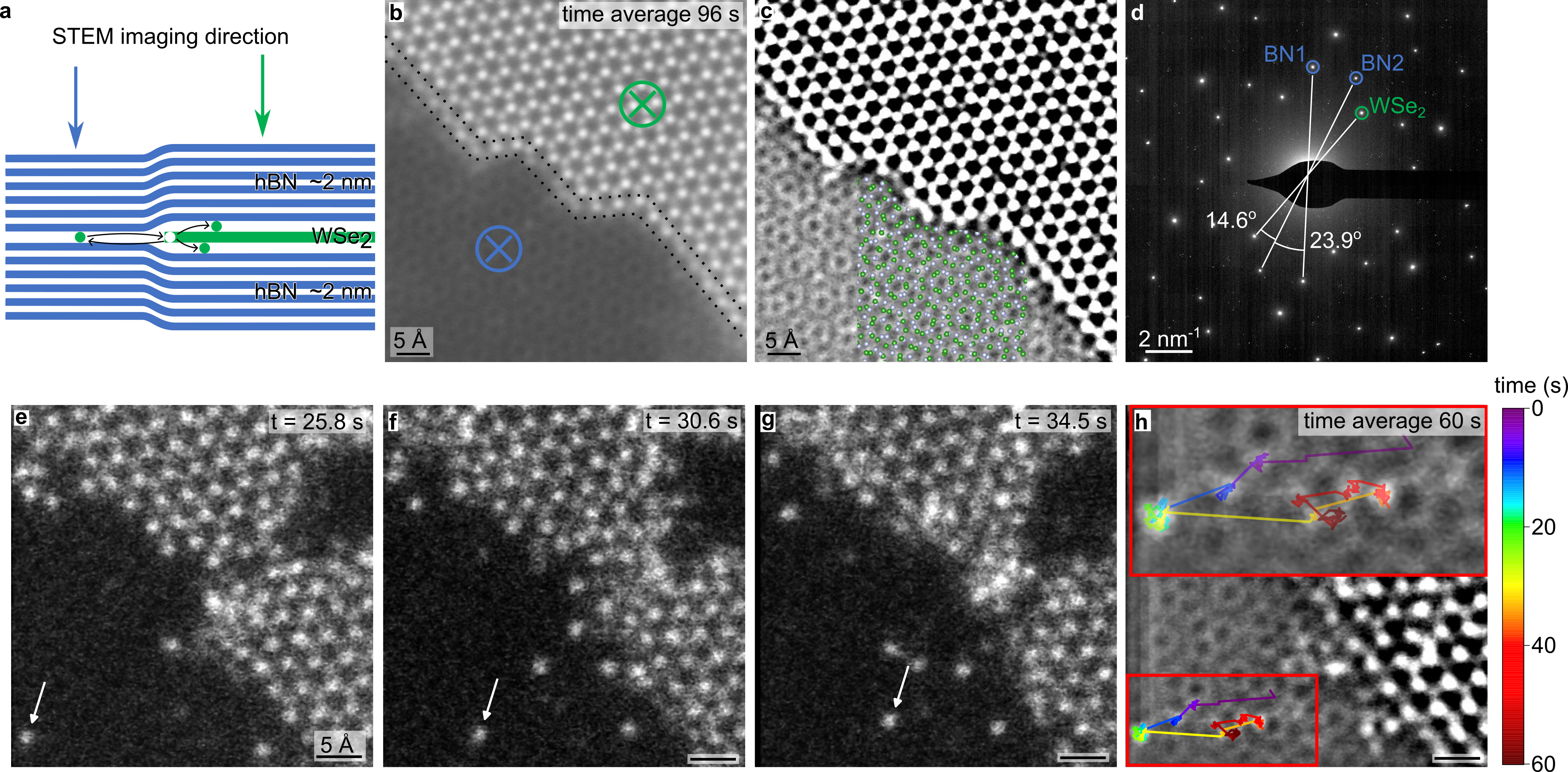}}
	\caption{\label{samplefig}
		\textbf{Sample overview.} 
		\textbf{(a)} Cross sectional sample schematic: Monolayer WSe$_{2}$ encapsulated within $\sim$2~nm thick BN on both sides.
		\textbf{(b)} Sum of over 400 drift corrected STEM-HAADF images at the WSe$_2$ edge from Supplementary Video 1. The blue and green arrow tails indicate the same positions as shown by the arrows in \textbf{(a)}. 
		\textbf{(c)} Same as \textbf{(b)} but with its own Gaussian blurred (15~pixel radius) version subtracted in order to emphasize the BN/BN moiré. An atomic model of twisted bilayer BN is shown on top of the BN/BN moiré.
		\textbf{(d)} Diffraction pattern of the sample. The blue and green circles indicate a $\{11\overline{2}0\}$ spot from BN and WSe$_{2}$, respectively. The white solid lines indicate the twist angles between the different layers. 
  		\textbf{(e-g)} Selected images from Supplementary Video 2. The white arrows point to an atom tracked at the BN/BN interface. Scale bars are 5 Å. 
    	\textbf{(h)} Time averaged image. The trajectory of the tracked atom is shown on top of the image. The inset shows an enlarged image of the trajectory. 
		}
\end{figure*}

Here, we directly visualize the atomic scale dynamics of W diffusion inside a vdW heterostructure, at the interfaces between BN/BN, BN/WSe$_2$, and BN/vacuum. We show that the diffusion properties are governed by defects at the interfaces, which are created continuously by the electron beam, and which create trapping sites for W atoms. This leads to diffusion properties that are more affected by defect densities than the details of the interface structure such as the size of the vdW gap. We show that it is possible to span the range from an intrinsic regime with very few defects to a highly defective regime where W atoms undergo small displacements between defect sites.

The ability to image the dynamics of atoms embedded inside vdW heterostructures with high spatial and temporal resolution that we demonstrate here offers opportunities for quantifying diffusion processes and observing atomic interactions directly. The precise identification of different atomic species and their beam-controlled motion may even be possible \cite{dyck2017placing, susi2022identifying}. 
Furthermore, our high-resolution imaging of individual atoms within BN-encapsulated heterostructures suggests that it may be possible to combine electrical and structural measurements of the active material in a vdW device through appropriate experimental design.

\section{Atomic Motion in the van der Waals Gap}
To study the atomic motion of W atoms at vdW-bonded interfaces, we fabricated BN/WSe$_2$/BN heterolayer stacks with monolayer WSe$_2$ and $\sim$2~nm thick BN, i.e., approximately 6 monolayers of BN. Details of sample fabrication are described in \textit{Methods}. By choosing WSe$_2$ crystals that were laterally smaller than the BN crystals, we embedded an edge of WSe$_2$ within the heterostructure. This served as the source of W (and Se) atoms, as discussed below. A schematic of this BN/WSe$_2$/BN heterostructure is shown in Fig.~\ref{samplefig}(a). Optical microscopy images of the samples are given in Figs.~S1 and S2. Such samples allow us to make quantitative comparisons between the diffusion of W atoms at the BN/WSe$_2$ and BN/BN interfaces. We measure diffusion at the BN/vacuum interface by creating simpler BN/WSe$_2$ heterostructures, consisting of $\sim$2~nm thick BN partially covered with monolayer WSe$_2$.

\begin{figure*}[t]
	\scalebox{\figurescale}{\includegraphics[width=1\linewidth]{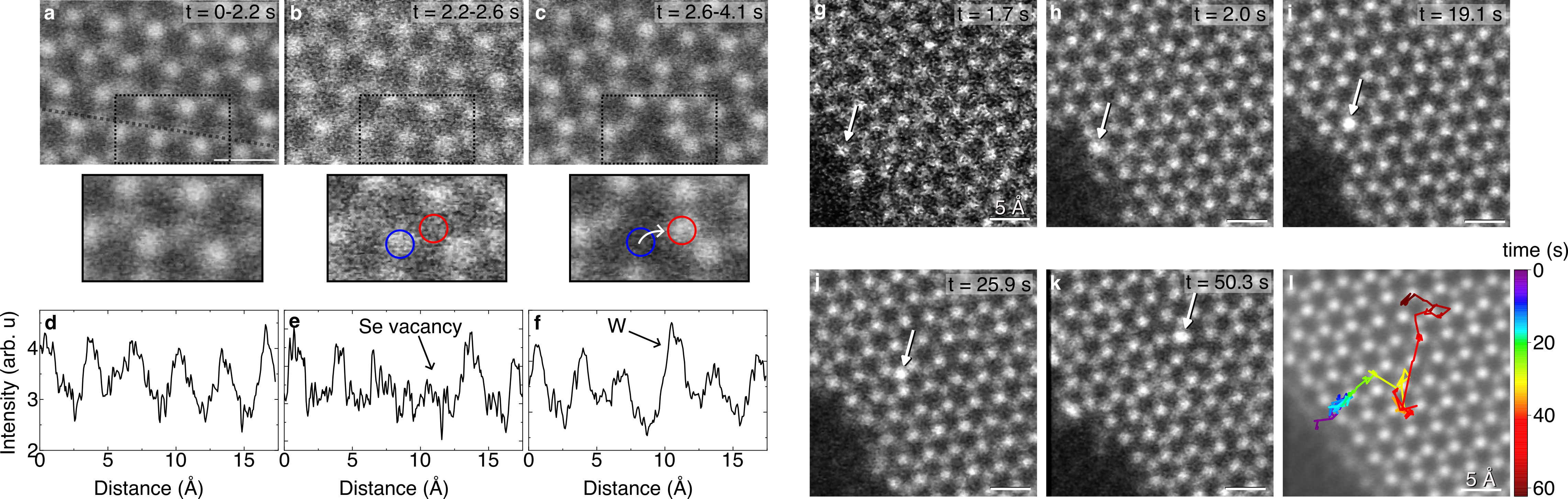}}
	\caption{\label{fig:Wremoval}
		\textbf{Creation of mobile W atoms.} 
		From the WSe$_{2}$ bulk:
        \textbf{(a--c)} Data from Supplementary Video 3 - STEM-HAADF images of a BN/WSe$_{2}$/BN heterostructure. The insets show the region in the dashed rectangle. 
		\textbf{(a)} Sum of 10 subsequent frames, t=0-2.2 s.
		\textbf{(b)} Sum of 2 frames, t=2.2-2.6 s. A Se vacancy has formed in the lattice site denoted by the red circle. The blue circle indicates the W atom that subsequently is ejected from the lattice.
		\textbf{(c)} Sum of 7 frames showing the indicated W atom has moved to the Se vacancy site, t=2.6-4.1 s.
		\textbf{(d--f)} Line profiles (5 pixel wide) along the dashed black line shown in (a).
        From the WSe$_2$ edge:
        \textbf{(g--k)} Data from Supplementary Video 4 - images of a tracked W atom (indicated by white arrows) that originated at the WSe$_2$ edge, and diffused at the BN/WSe$_2$ interface.
        \textbf{(l)} Time averaged image with the trajectory of the atom shown on top.
		}
\end{figure*}

Figure~1(a) showed the ideal situation with pristine interfaces. In reality, during assembly of vdW heterostructures it is possible to form bubbles between the layers with trapped hydrocarbon contamination \cite{haigh2012cross, kretinin2014electronic, schwartz2019chemical, pizzocchero2016hot, lee2022strong}. 
However, optical images of our heterostructure samples do not reveal visible bubbles on the regions containing monolayer WSe$_2$. Some bubbles in regions with thicker BN are indicated in Fig.~S1 and S2, but these areas were not used for our experiments. The contamination within bubbles would in any case be seen in STEM or TEM imaging due to the added mass of the hydrocarbon, but are not visible in overview images of the samples (Fig.~S3). Bubble-free regions typically form clean interfaces due to a squeeze-out effect that pushes contaminants away \cite{haigh2012cross, pizzocchero2016hot}, especially when prepared with dry transfer methods as used here. However, the interlayer gap may differ from the gap predicted from DFT calculations if either material is susceptible to oxidation \cite{rooney2017observing}. 
In particular, the BN/WSe$_2$ interlayer gap was found to be slightly larger than expectations from DFT calculations \cite{rooney2017observing}. Since we also prepared samples at ambient conditions we would expect a similar interlayer spacing.  
 
In our samples, the sensitivity of HAADF-STEM for heavier atoms allows us to image the monolayer WSe$_2$ and individual diffusing W atoms at sub-0.1~nm resolution through the BN layers (Fig.~\ref{samplefig}(b, c), Supplementary Video 1). Although individual Se atoms were also mobile and visible within the sample, we tracked the motion of W atoms because they appear brighter in HAADF-STEM imaging, as contrast scales with atomic mass as Z$^{1.7}$ \cite{hartel1996conditions}. The layers were not rotationally aligned during the stacking process, and the rotational misalignment of the three layers can be measured from diffraction, as in Fig.~\ref{samplefig}(d). We measure a rotation angle of 23.9$^{\circ}$ between the two BN crystals. In Fig.~\ref{samplefig}(c) we have superimposed an atomic model of twisted bilayer BN rotated to 23.9$^{\circ}$ which matches with features in the image. We use fixed conditions (electron beam energy and current, dwell time, magnification, and number of pixels per image) for all data used to quantify diffusion dynamics (see \textit{Methods} for imaging conditions and data analysis). All data shown in the main text is obtained with an electron beam energy of 200~keV and we provide additional data obtained at 60~keV in the Supplementary Information. 

During imaging of these structures we then locate mobile W atoms that originate from the WSe$_2$ lattice. They can either diffuse at the interface between the two BN crystals or at the interface between the WSe$_2$ and BN. Figure~\ref{samplefig}(e-g) shows images of a W atom diffusing at the BN/BN interface, see also Supplementary Video 2. Figure~\ref{samplefig}(h) is an averaged image from the video with the trajectory of the W atom also shown. From such trajectories we can study the atomic scale dynamics and statistics or the W atom motion at the different interfaces. These videos show atomic trajectories that resemble a diffusion process with intermittent trapping (Fig.~1(h)) \cite{skaug2013intermittent}, where an atom makes displacements between distinct sites at which it resides for some time. We note that even with a frame time of 0.217 s, atoms remain at such sites for multiple frames. Thus the observations can capture the dynamics of atomic residence at certain sites, even if diffusion processes occur on a faster time scale where rapid jumps are likely not resolved.

In the following sections we describe and analyze the formation of diffusing W atoms and W diffusion process at the atomic scale at the WSe$_2$/BN, BN/BN, and BN/vacuum interfaces.

\begin{figure*}[t]
	\scalebox{\figurescale}{\includegraphics[width=1\linewidth]{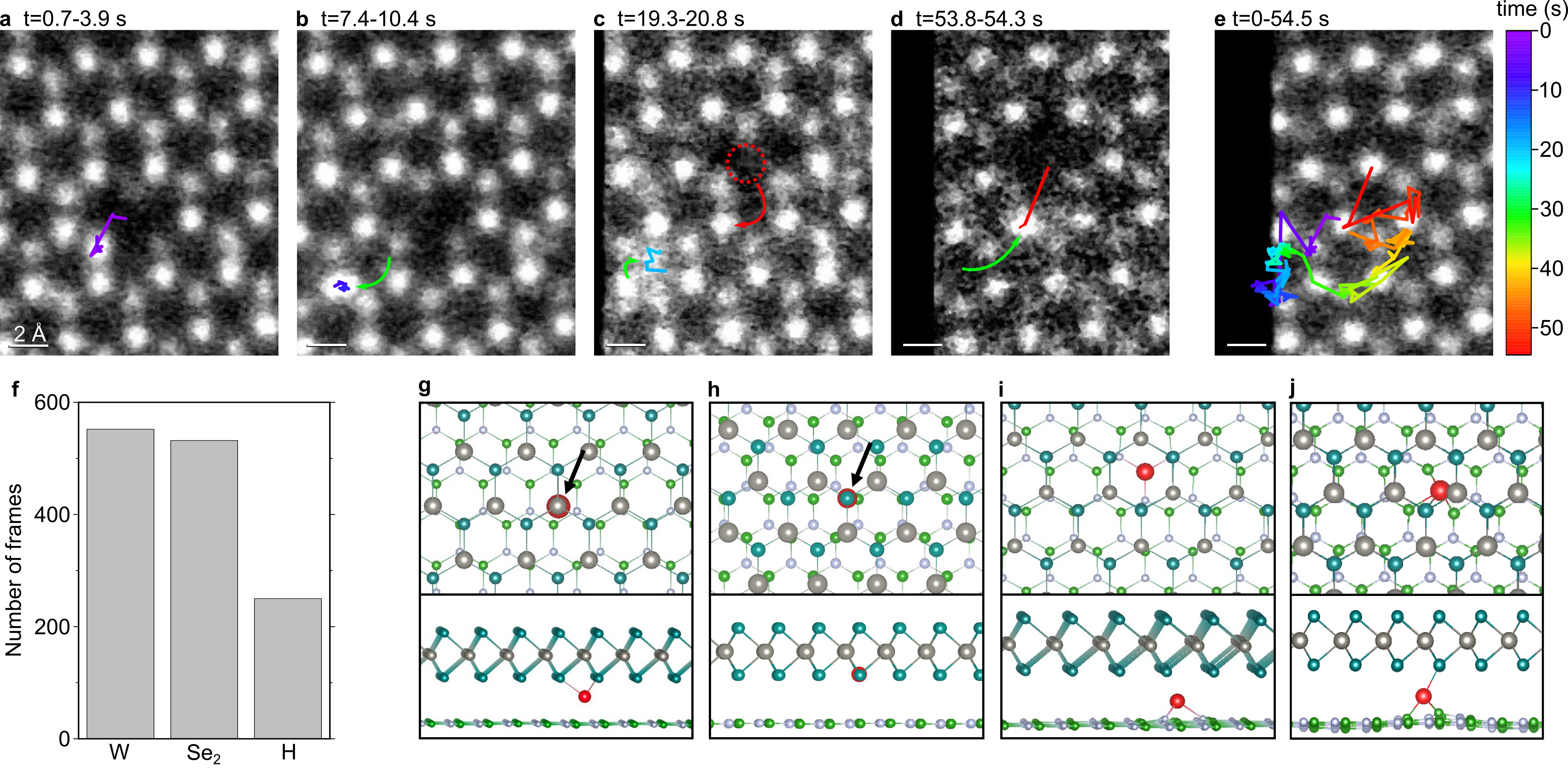}}
	\caption{\label{fig:diff-BN-WSe2-v2}
		\textbf{Atomic motion in BN-WSe$_{2}$ interface.} \textbf{(a--d)} STEM-HAADF images obtained from Supplementary Video 5 by summing the frames corresponding to the time interval indicated above each panel.
		\textbf{(a)}, \textbf{(b)}, and \textbf{(c)} show snippets of the full trajectory, with the atom residing on a Se$_2$, W, and hollow site, respectively. Tracks from the diffusing atom are overlaid to the images. In (\textbf{a)} the trajectory for t=0--3.9~s also shows the origin of the atom. In \textbf{(d)} the trajectory for t=53.8--54.5~s also includes the last recorded position where the atom reincorporates into the lattice. 
		\textbf{(e)} STEM-HAADF image obtained by summing all frames in the video with tracks of the atom overlaid.
		\textbf{(f)} Statistics of W resting sites obtained from the data shown in this figure as well as Fig.\ S4 and S5.
		\textbf{(g-j)} DFT-relaxed atomic structures. For computational simplicity we have chosen to calculate a rotationally aligned BN/WSe$_2$ heterostructure, see \textit{Methods} for details. 
        \textbf{(g)} The most stable position for W in a pristine BN/WSe$_2$ heterostructure. 
        \textbf{(h)} The most stable position for W near a Se vacancy. 
        \textbf{(i, j)} Stable positions for W in the presence of a B and N vacancy, respectively. The different positions of W with respect to the WSe$_2$ layer illustrates that vacancies in the BN can greatly affect the preferred location of the W atom.
		}
\end{figure*}

\section{Diffusion at the BN/WSe$_{2}$ Interface}
We first discuss the origin of the moving W atoms at the BN/WSe$_{2}$ interface. Upon initially imaging a pristine BN/WSe$_{2}$/BN region, we observe that W ejection from the bulk WSe$_{2}$ is preceded consistently by the formation of neighboring single or double selenium vacancies (Fig.~\ref{fig:Wremoval}(a-c), Supplementary Video 3). Another example is given in Fig.~S4. This takes place on a timescale of approximately 1 min. The knock-on threshold for W atoms in a pristine WSe$_{2}$ lattice is above 500 keV while the threshold for Se atoms is 190 keV \cite{komsa2012two}. The observation of Se vacancy formation as the first step of the process is therefore consistent with a beam-driven effect. 

We have also observed W atoms at the edge of the WSe$_2$ lattice diffuse into the BN/WSe$_2$ interface (Supplementary Video 4). Figure~\ref{fig:Wremoval}(g-k) shows images with a moving W atom indicated by a white arrow. Figure~\ref{fig:Wremoval}(l) shows the trajectory of the atom.

After their release from the WSe$_{2}$ lattice we track the diffusion behavior of single W atoms at the BN/WSe$_{2}$ interface. Figure~\ref{fig:diff-BN-WSe2-v2}(a-d) and Supplementary Video 5 follows a W atom as it is ejected from the WSe$_{2}$ lattice and sits at different times in a Se$_{2}$ site (Fig.~\ref{fig:diff-BN-WSe2-v2}(a)), a W site (Fig.~\ref{fig:diff-BN-WSe2-v2}(b)), and a hexagonal or “hollow” (H) site (Fig.~\ref{fig:diff-BN-WSe2-v2}(c)). It finally becomes reincorporated into the WSe$_{2}$ lattice (Fig.~\ref{fig:diff-BN-WSe2-v2}(d)), albeit at a different location from its origin because the W vacancy diffused a distance of one unit cell, as shown in Fig.~\ref{fig:diff-BN-WSe2-v2}(c). Since the W atom can reincorporate into the WSe$_{2}$ lattice, this suggests that the W atoms diffuse directly at the BN/WSe$_{2}$ interface, and do not diffuse into another vdW gap. Figure~\ref{fig:diff-BN-WSe2-v2}(e) displays the total trajectory. 
See Fig.\ S5 and S6 for trajectories of additional W atoms measured at the BN/WSe$_{2}$ interface. The data shown in Fig.~\ref{fig:diff-BN-WSe2-v2}, Fig.\ S5, and S6 are consistent in that the W atoms reside on W and Se$_{2}$ sites at about the same rate, 41\% and 40\% of frames, respectively, while they reside on a H site on 19\% of frames (Fig.~\ref{fig:diff-BN-WSe2-v2}(f), see \textit{Methods} for details).

\begin{figure*}[]
	\scalebox{\figurescale}{\includegraphics[width=1\linewidth]{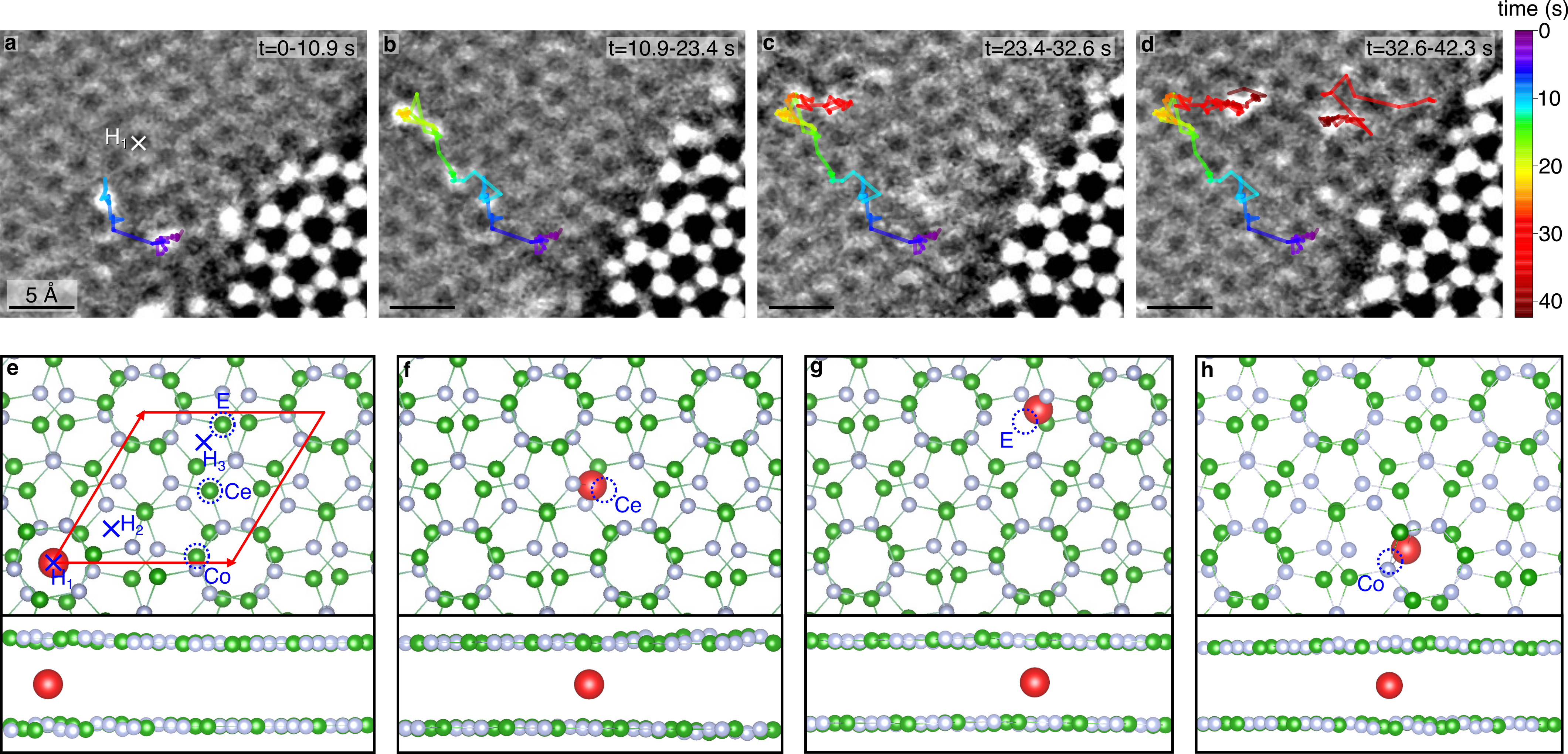}}
	\caption{\label{fig:diff-BN-BN}
		\textbf{Atomic motion in BN-BN interface.} 
		\textbf{(a--d)} Atomic trajectories recorded in the time intervals t=0--10.9, 10.9--23.4, 23.4--32.6, and 32.6--42.3 s, respectively. The trajectories for previous intervals are included in later ones (data from Supplementary Video 6). In (d) another W atom emerges from the WSe$_2$ edge.
		\textbf{(e--h)} DFT-relaxed atomic structures of a W atom intercalated between two BN layers. For simplicity this simulation only includes one layer on either side of the interface. Green, silver, and red spheres denote the B, N, and W atoms. (e) A pristine BN/BN interface with characteristic hollow and lattice sites indicated in blue. The red arrows indicate the moiré unit cell (see \textit{Methods} for the construction). In a pristine BN bilayer, the W atom tends to sit at H$_1$. 
        \textbf{(f--h)} W atom stable sites in the presence of a Ce (f), E (g), or Co (h) defect. All three configurations show that sites near the B vacancy are preferred over H$_1$. 
		}
\end{figure*}

We next compare the statistics in Fig.~\ref{fig:diff-BN-WSe2-v2}(f) with expectations from DFT calculations (\textit{Methods}) to better understand the potential landscape experienced by the interstitial W atoms. We perform calculations of the BN/WSe$_2$
interface, for simplicity including only monolayer BN in the calculation. We do not expect the results to change if we included multiple BN layers, based on our calculations for the free surface of BN of different thicknesses (see \textit{Diffusion at the BN/Vacuum Interface}). We find that the most stable site for an interstitial W atom in a pristine WSe$_{2}$/BN interface is directly beneath another W site (Fig.~\ref{fig:diff-BN-WSe2-v2}(g)). 

We then computationally introduce vacancies into the system, adding both N and B vacancies since both vacancy types could be present: B vacancies were considered more common during TEM imaging of BN \cite{kotakoski2010electron, zhang2020point}, although recent literature suggests that both the accelerating voltage and vacuum level could change the prevalence. Indeed, at 200 keV, N vacancies may be more prevalent than B vacancies \cite{bui2023creation}. Unsurprisingly, Se, B, and N vacancies (Fig.~\ref{fig:diff-BN-WSe2-v2}(h, i, j)), respectively) are stable sites for the W atom, as it can form bonds to the three neighboring W, N, or B atoms, respectively. This does not depend on the location of the vacancy with respect to the moiré unit cell, as shown in Fig.~S7(a), where a W atom behaves similarly when the B vacancy is in a different location. Furthermore, DFT calculations indicate that B and N vacancies behave in a qualitatively similar manner. Since B and N sites can not be distinguished in our experimental images, we discuss them together in the following analysis.

The interlayer distance appears to have only a small effect on the equilibrium position. In Fig.~S7(b) we compare the equilibrium position of a W atom in a N vacancy for the DFT-predicted interlayer spacing (as used in Fig.~3(j)), and an experimentally determined interlayer spacing which is slightly larger \cite{rooney2017observing}. The equilibrium positions differ only slightly (see \textit{Methods} for details) and we do not expect this small effect to  impact our classification of resting sites given in Fig.~3(f).

If vacancies in the BN were to completely determine W pinning sites, then we would expect W atoms to reside equally on W, Se$_{2}$, and H sites since vacancies in the BN layers would be distributed above these sites equally. Since we find that W atoms reside mainly on W and Se sites, this indicates that the potential landscape of WSe$_{2}$ also plays a role in determining pinning sites. This is supported by previous calculations of Au diffusion in pristine heterostructures of graphene and transition metal dichalcogenides, where the calculated potential landscape predicted that diffusion pathways are along the W and Se$_2$ sites \cite{iyikanat2014ag}. In our data, the equal prevalence of W and Se sites might further be due to the presence of Se vacancies, because in the pristine case the W site is preferred. Hence, analysis of the preferential resting sites for these interstitial atoms reveal that their motion is influenced by the presence of Se vacancies in the WSe$_{2}$ as well as B or N vacancies in the top and bottom BN layer. In the following we will show that W diffusion at the BN/BN interface exhibits analogous characteristics. 

\begin{figure*}[t]
	\scalebox{\figurescale}{\includegraphics[width=1\linewidth]{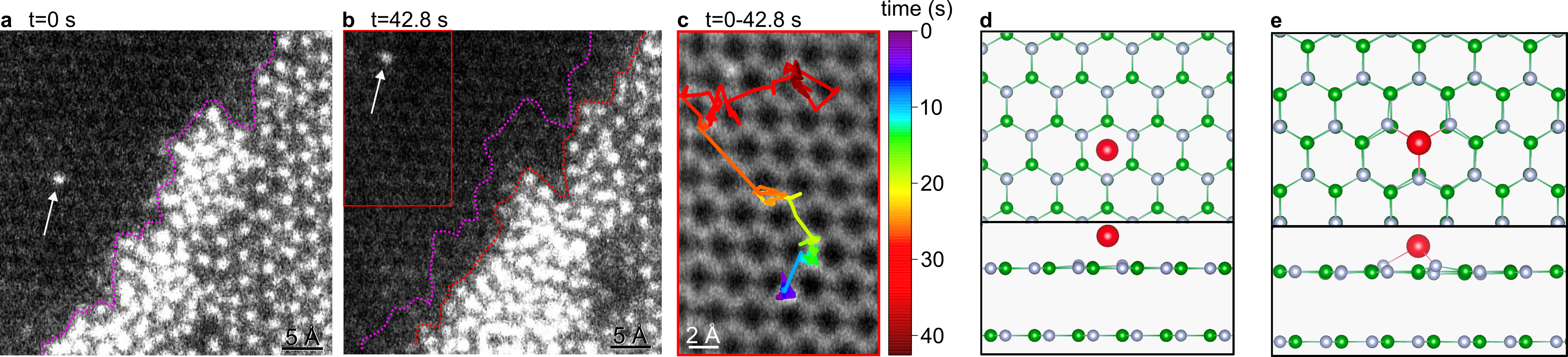}}
	\caption{\label{fig:BN/vac}
		\textbf{
		W diffusion at the BN/vacuum interface.
		} 
		\textbf{(a, b)} First and last frame, respectively, from Supplementary Video 7, showing a W atom on the surface of $\sim$2~nm BN. The purple and red dotted line indicates the edge of the WSe$_{2}$ layer at the start and end of the video. The tracked atom is indicated with a white arrow. 
		\textbf{(c)} STEM-HAADF obtained by summing 197 frames of the video. The trajectory of the W atom is shown on top.
		\textbf{(d)} DFT-relaxed structure showing the most stable position for a W atom on a pristine BN bilayer.
		\textbf{(e)} DFT relaxed structure of W near a B vacancy on a BN surface. For computational efficiency the BN in the calculation is two monolayers thick.
		}
\end{figure*}

\section{Diffusion at the BN/BN Interface}
Upon imaging a pristine BN/BN region near a WSe$_{2}$ edge, we find several mobile W atoms. We assume that these atoms are detached from the WSe$_{2}$ edge through interactions with the 200 kV electron beam because undercoordinated edge atoms are easier to eject. We find that the edge does not retract noticeably over extended periods of time (Fig.~\ref{samplefig}(b)), but edge atoms detach and reattach analogous to a sublimation and deposition process on a surface. Only the outer row of atoms on the WSe$_{2}$ edge (highlighted in Fig.~\ref{samplefig}(b)) appears blurry in the time averaged-image (over 96 s) due to their motion. 

To quantify the atomic scale details of W diffusion in the BN/BN interface, Fig.~\ref{samplefig}(e-h) and  Fig.~\ref{fig:diff-BN-BN}(a-d) show trajectories of W atoms at high magnification. See Fig.~S8 and S9 for trajectories of additional W atoms at the BN/BN interface. The atoms make several smaller displacements, $<$1~Å, seemingly around pinning sites, and fewer, larger displacements between such sites. 
We observe that the W atoms reside mainly on the moiré lattice of the BN/BN interface, i.e.\ on BN lattice sites. We compare this observation to DFT calculations for stable resting sites at the BN/BN interface. Figure~\ref{fig:diff-BN-BN}(e) shows the model of the BN/BN interface that we used for these calculations (\textit{Methods}). The moiré unit cell features some distinctive sites that are indicated with blue circles and crosses in Fig.~\ref{fig:diff-BN-BN}(e). The largest hollow space, denoted H$_1$, is present at the corners of the unit cell, while smaller hollow spaces are denoted by H$_2$ and H$_3$. Figure~\ref{fig:diff-BN-BN}(e) shows that for a pristine BN/BN interface, the most stable site for the W atom is at H$_1$. This corresponds to the center of the “donut” shapes in Fig.~\ref{fig:diff-BN-BN}(d) (marked with a white cross). Configurations with W atoms at H$_2$ and H$_3$ display about 200~meV higher energy than a configuration with a W atom at H$_1$ (Fig.~S7(c, d)). However, introducing B vacancies at three distinct lattice sites on the moiré unit cell, namely the center (Ce), edge (E), and corner (Co) sites (Fig.~\ref{fig:diff-BN-BN}(f-h)) makes these sites significantly more stable positions for the W atoms. For example, in the case of an E site, the DFT calculations show that the total energy of the system increases by 7~eV when moving a W atom from E to H$_1$, see Fig.~S7(e, f), because 3 W-N bonds must be broken. This means that B vacancy sites are significantly more stable sites for W interstitials as compared to the ‘best’ H$_1$ site in the pristine interface.

Overall, DFT simulations predict that the H$_1$ site should be the most stable resting site for W atoms in a pristine interface. However, we find experimentally that W atoms reside on the BN/BN moiré lattice in agreement with our DFT predictions that vacancies in the Ce, E, or Co sites are significantly more stable sites. These results therefore suggest that defects formed by the electron beam govern resting sites and diffusion properties at the BN/BN interface. This behavior is consistent with that seen for the BN/WSe$_{2}$ interface. 

\section{Diffusion at the BN/Vacuum Interface}
We finally image W atoms on the free surface of $\sim$2~nm thick BN for comparison with the situations described above where the W atoms are encapsulated. For this experiment, we fabricate samples where monolayer WSe$_2$ is placed on a $\sim$2~nm thick BN crystal which is wider laterally than the WSe$_2$ crystal. Upon imaging the WSe$_2$ edge, we find several W atoms which have been released from the WSe$_2$ lattice and diffuse on the hBN surface.

Figure~\ref{fig:BN/vac}(a,~b) display the first and last image, respectively, from a video where we track one of these atoms. These images reveal that the WSe$_2$ edge recedes over time, with dislodged W/Se atoms aggregating in a structure on the edge that appears three dimensional, based on the proximity of the W atoms in projection. This behavior differs from the BN encapsulated WSe$_2$ edge, where spatial confinement evidently prevents atoms on the WSe$_2$ edge from forming a 3D structure, so they instead detach and reattach in a 2D fashion (Fig.~\ref{samplefig}(a)). Figure~\ref{fig:BN/vac}(c) shows an average of the entire video with the atom trajectory shown on top. We find that the atom makes several smaller displacements less than approximately an atomic spacing around a pinning site and fewer larger displacements over several atomic spacings in between such sites. Furthermore, we see that the pinning sites coincide with positions on the BN lattice. 

Before performing quantitative analysis of the observed motion, it is important to establish whether the W atoms in experiments such as that in Fig.~5 are moving on the surface of the BN or between layers, as in Fig.~4. Although we cannot measure the z-direction position from the images, several factors suggest the W atoms are on the surface. First, W atoms are released on the BN surface and the energetics of diffusion through a BN layer and into a vdW gap are unfavorable, even in the presence of beam-induced defects. 
Second, the analysis to be shown below yields consistent and different values for the diffusion coefficient in these experiments compared to those at the BN/BN interface; we attribute this to the presence of surface contamination \cite{zan2011metal}. 
Finally, in data such as Fig.~3(a-e) a W atom is released from the WSe$_2$ lattice and later reincorporated. This particular atom therefore remained in the gap where it was released, showing that W atoms can make extended
movements without changing planes.

Figure~\ref{fig:BN/vac}(d) shows results from DFT calculations that predict the most stable site for W atoms on a pristine BN surface is in the hollow site, as minimal orbital overlap gives the maximum spacing for the adatom. B vacancies are a far more stable site for W adatoms (Fig.~\ref{fig:BN/vac}(e)). Our experimental observation that W atoms occupy lattice sites on the BN is consistent with the importance of defects in the BN layer in determining pinning sites. 

In these calculations, the thickness of BN does not affect the results. This is shown in Fig.~S7(g, h) where we find a similar stable site for a W atom on a pristine surface and at a B vacancy, whether the BN thickness is monolayer or bilayer. Furthermore, N vacancies behave similarly to B. Figure~S7(i) shows that N vacancies lead to stable sites with even higher binding energy (we calculate defect formation energies of 6.75 eV and 9.65 eV for W in a B or N vacancy, respectively, on monolayer BN; details in \textit{Methods}). This suggests that W atoms are more likely to bind to a B vacancy compared to an N vacancy at equilibrium conditions although local pertubations may alter the situation. 
While we cannot identify a specific lattice site on the BN surface as N or B, experimentally we also do not observe a preference for either of the B or N lattice sites.

\begin{figure*}[t]
	\scalebox{\figurescale}{\includegraphics[width=1\linewidth]{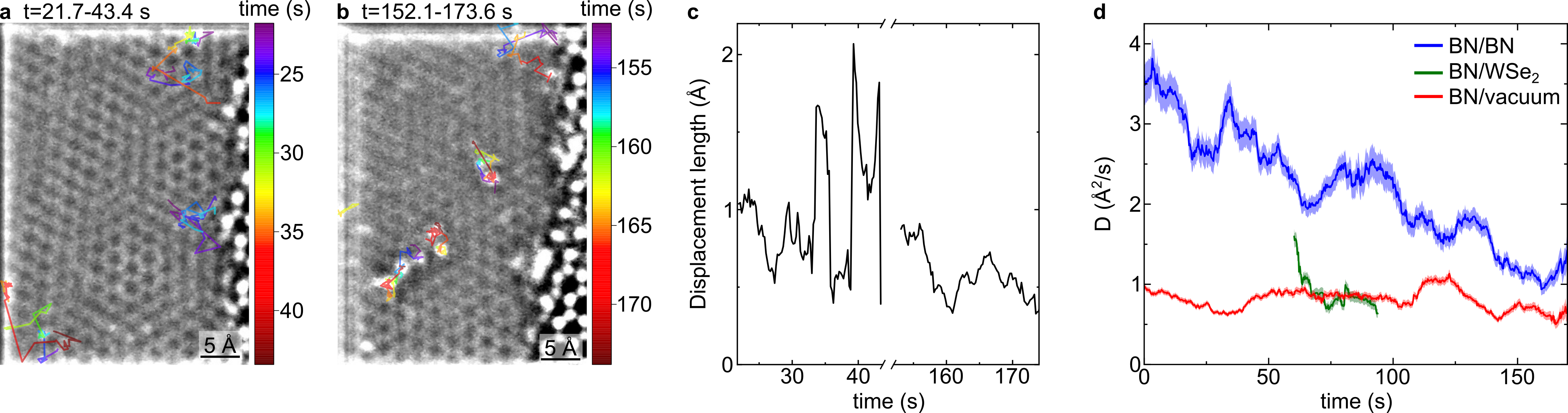}}
	\caption{\label{fig:MSD}
		\textbf{Time-dependent diffusion properties.} 
        \textbf{(a)} All atomic trajectories in a 21.7~s time interval (corresponding to 100 images) starting at $t$=21.7~s. 
        \textbf{(b)} All atomic trajectories in a 21.7~s time interval starting at $t$=152.1~s. 
        \textbf{(c)} A running average of the displacement length of the atomic trajectories shown in (b, c). The running average is constructed by, first, averaging the displacement length of all atoms in a given time step, and, second, performing a running average over 10 time steps (2.17~s). The larger spikes at, e.g. about t=34 and 40~s are due to the large displacements at occurring at these times (see panel (b)). 
        \textbf{(d)} The W diffusion coefficient averaged across all tracked atoms and plotted against time for the BN/BN, BN/vacuum, and BN/WSe$_2$ interface, respectively. The diffusion coefficient was calculated using a co-variance-based estimator \cite{vestergaard2014optimal} on a rolling time-window of 15~s. The time used for plotting is the start of each window. The shaded areas in each plot correspond to the error bars. We note that the data from the BN/WSe$_2$ interface is obtained by manual tracking of atoms since automatic tracking of an extra W atom on a background lattice of WSe$_2$ is challenging. See \textit{Methods} for details.  
		}
\end{figure*}

\section{Statistics of the Atomic Motion at Each Interface}
Having shown qualitatively that W atoms appear to move through displacements between defect sites, we focus now on a quantitative analysis of the time-dependence of this behavior. This quantitative analysis simplifies considerably if the diffusion coefficient does not depend on direction. Therefore, we first investigated the diffusion isotropy, see Fig.~S10--S13. We found that the diffusion is isotropic except for the shortest displacements, less than one atomic spacing. These display some degree of anisotropy aligned with the axes of the image, implying that the result is due to measurement artifacts (see caption Fig.~S10). We further note that we do not find any evidence that the motion is affected by the electron beam scanning direction. 

Since defects are continuously produced by the electron beam during the experiment, we expect a gradual increase in the frequency with which diffusing atoms are trapped on defects. This lowers the diffusion coefficient over time. To test this hypothesis, we next investigate the time-dependent diffusion coefficient. 

We performed experiments where we acquired data for $\sim$170~s from approximately the same area of a sample, by minimizing sample drift, and measured the diffusivity of the atoms in these videos. For this data, we define t=0 as the time where a pristine area on the sample is first imaged. For the BN/BN interface we observe a marked decrease in W atom diffusivity. This is visible, for example, in data such as Supplementary Video 8, which was recorded from t=0. Such a trend is not apparent for the BN/WSe$_2$ and BN/vacuum interfaces as we will show and discuss below. 

At the start of an experiment, tracking single atoms is challenging because their displacements between video frames can be large before many defects have been formed by the electron beam (Fig.~S14). Shortly afterwards, 
Fig.~6(a) shows W atom trajectories from early times (100 frames, corresponding to 21.7~s) of an experiment while Fig.~6(a) shows W atom trajectories from the last 21.7~s from the same experiment. We find that their average displacement length decreases over time (Fig.~6(c)). 

We then average across all trajectories recorded to find the overall behavior of W atoms at each interface. The diffusion coefficient, $D$, is optimally estimated using a co-variance based estimator \cite{vestergaard2014optimal} (\textit{Methods}). We used this method on all displacements observed within a rolling time-window of 15~s to elucidate the time-dependence of the diffusion coefficient.

Figure~6(d) shows the W diffusion coefficient plotted against time for each interface. For the BN/BN interface, $D$ decreases over time, starting at a value of 3.5~Å$^2$/s at t=0~s and ending at a value of 1.5~Å$^2$/s at t=150~s. We expect the initial value of 3.5~Å$^2$/s is a lower bound of $D$ because we find that the W atoms initially undergo large displacements which are challenging to track. When several atoms are present near each other this would result in an underestimation of their displacements with the nearest-neighbor tracking algorithm (see \textit{Methods}) we used.  

Imaging beyond approx.\ 150~s results in the loss of BN/BN moiré image intensity, see Fig.\ S15-S17. This indicates damage in the BN structure, and hence is consistent with the hypothesis that the decreasing diffusion coefficient is due to the continuous introduction of defects until the electron beam drills a hole through the sample with W atoms agglomerated along the edges of the hole - see in particular Fig.~S15 and S16. Such an interpretation is supported by literature. It is known that electron beam generated defects are not created uniformly through the thickness of a 2D material, but form more readily at the bottom surface \cite{kotakoski2010electron}. TEM experiments in literature with BN have also shown that BN tends to be removed in a layer-by-layer fashion from the bottom surface with $\sim$6$\cdot 10^6$~electrons/Å$^2$ needed for the removal of a layer \cite{gilbert2017fabrication}. We use a beam current of 50 pA to scan a 4$\times$4~nm area, giving a dose rate of 2$\cdot 10^5$ electrons/Å$^2$/s. Thus, we need $\sim$30~s to remove a BN monolayer, fitting well with our observation of 150~s needed to etch through 2~nm of BN, corresponding to 6~monolayers of BN. 

In addition, experiments at 60 kV show that degradation of the BN/BN interface and formation of W vacancies in bulk WSe$_2$ occurs at significantly longer time scales. At 60 kV, imaging beyond 400-600~s is needed before any noticeable BN/BN moiré intensity loss (Fig.~S18 and S19, Supplementary Video~9). Imaging the BN/WSe$_2$ interface, a pristine region was imaged for almost 400~s before the formation of a W vacancy (Fig.~S20, Supplementary Video~10). Thus, defect formation is highly dependent on electron beam energy, and use of lower electron beam energies may be a viable strategy for observing a sample without significant defect formation or modification.  

At the BN/vacuum interface, the diffusion coefficient for W atoms fluctuates around a value of 0.8~Å$^2$/s. In literature, the diffusivity of intercalated atoms in vdW materials has been shown to increase with interlayer distance \cite{yoo2017fast, rasamani2017interlayer}. Hence, intuitively one would expect that atoms diffusing on the free surface of BN would display faster diffusion compared to atoms diffusing in a vdW-bonded interface. 
We believe that two factors may account for the lower diffusivity at the BN/vacuum interface compared to the BN/BN interface. First, polymeric contamination is ubiquitous on the surface of 2D materials samples that have been transferred to TEM grids using polymer handles \cite{lin2012graphene}. Such carbon-containing contamination is challenging to image through a multilayer due to its low atomic mass and is known to pin atoms diffusing on the surface of 2D materials \cite{zan2011metal}. Second, the data from the BN/vacuum interface comes from a sample where W atoms diffused on the bottom side of the BN with respect to electron beam direction. The bottom surface of a 2D material is more prone to electron beam induced defects, since atoms can be removed more easily by direct knock-on damage \cite{komsa2012two}. Thus, the surprising result that the diffusion coefficient of W atoms at the BN/vacuum interface is significantly smaller than that of W atoms at the BN/BN interface suggests that the diffusion properties are less affected by interlayer distances, and more strongly depend on the defect concentration and the presence of contamination on the sample.

Finally, for the BN/WSe$_2$ interface, we find that W atoms are released from the WSe$_2$ lattice typically after at least 60~s of imaging. The start of each recorded trajectory is therefore set to t=60~s. $D$ appears to be decreasing initially and then fluctuates around a value of $\sim$0.8~Å$^2$/s, the same as the diffusion coefficient of W atoms at the BN/vacuum interface. This may be the result of the amount of imaging typically needed to release W atoms from bulk WSe$_2$, where by the time W is released the interface is already markedly damaged. 

Combining information for all three interfaces, we conclude that the comparative $D$ values and their variation with time are consistent with atomic motion dominated by impurities and defects that are created by the electron beam at the interface.

\section{Conclusions}
We have quantified diffusion of W atoms at the BN/vacuum, BN/BN, and BN/WSe$_2$ interfaces with atomic resolution imaging of single atoms through several atomic layers of BN, measuring not only diffusion parameters but also atomistic diffusion pathways. Our results show that the diffusion properties of W atoms in vdW bonded interfaces are tunable and highly affected by defects at the interface. We determine this by directly visualizing trapping sites during the diffusion and comparing with expectations from DFT calculations. We further quantify diffusion properties and find a strong impact of defect concentration on the diffusion coefficient of W atoms at the BN/BN interface. This results in a time-dependent diffusion coefficient as a result of defects added during STEM imaging at 200~kV acceleration voltage. In addition, we have performed experiments with a 60~kV acceleration voltage showing the formation of defects at a significantly slower rate. The strong dependence on defects suggests that the vdW crystal quality is very important to consider when designing intercalation and diffusion processes.  

We anticipate that this work will offer strategies for visualizing diffusion and atomic interactions within vdW gaps. Any relatively heavy atom, e.g., Ta, W, Pt, etc., can be imaged, and interactions between atoms with particular properties such as magnetic can be examined. It is also possible to perform the experiments in a TEM sample holder capable of heating or cooling the sample, to enable measurements say above and below a magnetic critical temperature. The vdW heterostructures we have examined, with their relatively thick insulating layers of $\sim$2~nm BN, are compatible with electrical property measurements, they permit simultaneously measuring atomic structure and electrical properties. 
Having showed that imaging with lower electron beam energy significantly reduces defect formation, the use of controlled electron beam dosing and low energy electrons will enable small modifications to be made and changes to electrical properties measured.

In addition, the use of other 2D materials for encapsulation, especially if composed of light elements, allows further possibilities. A conductive material like graphene enables quantification of the diffusion behavior as a function of electrical current and possibly atomic-level measurements of electromigration. Use of an anisotropic material like phosphorene enables studying directional diffusion. Using 2D materials with specific defect types, such as grain boundaries, allows directly evaluating the effect of such defects on diffusion and trapping. Finally, we highlight that novel TEM methods could aid in such experiments. Progress in image de-noising and machine learning for atom tracking could facilitate experiments at high frame-rates or low electron energies \cite{ede2021deep}, and electron beam programming and automated experiments could be used to create specific defects at predetermined sites \cite{kalinin2023machine}. This includes the possibility of engineering arrays of atoms by seeding defects in determined positions. The ability to image individual atoms with good time and space resolution therefore offers exciting prospects for future progress in atomic-level material design.

\section{Methods}

\footnotesize{

\subsection{Sample Fabrication}
WSe$_2$ crystals were purchased from HQ Graphene. BN and WSe$_2$ crystals were exfoliated onto an SiO$_2$ (90~nm)/Si substrate and suitable flakes were identified using optical contrast. The vdW heterostructures were fabricated using a sequential pick up of the top BN and monolayer WSe$_2$, and releasing this BN/WSe$_2$ heterostructure on the bottom BN. We used a poly(bisphenol A carbonate) (PC)-film-covered polydimethylsiloxane (PDMS) stamp on a glass slide for the transfers. Finally, the heterostructures were transferred to SiN/Si TEM grids, purchased from Norcada, Canada, using wedging transfer \cite{thomsen2017suppression}.

\subsection{STEM}
HAADF-STEM imaging was performed in a Titan Themis Z G3 Cs-corrected S/TEM with a beam current of 50~pA. All data in the main text was obtained with an acceleration voltage of 200~kV and a frame time of 0.217~s (512x512~pixels with a dwell time of 500~ns). In Fig.~S17-S19 we show data obtained with beam current of 50~pA, acceleration voltage of 60~kV, images with 512x12~pixels, and dwell times as listed in the captions. The Supplementary Information contains details on all Supplementary Videos (video length, sample type, scale). Raw data for all Supplementary Videos can be downloaded at \cite{thomsen2024atomic}.

\subsection{Atom Tracking}
Raw videos were drift corrected using a rigid registration tracking scheme \cite{NSscript} using ImageJ. Subsequently, single atoms were tracked using TrackMate \cite{ershov2022trackmate} which is a plugin to ImageJ. We used the nearest-neighbor tracker which links the nearest atoms in subsequent images, with a maximum linking distance of 1~nm. 

If there are several atoms present within the maximum linking distance this can create false linking between atoms. To minimize these events we exclude areas in our data with a large density of atoms and/or perform additional manual filtering of tracks to eliminate false links. We show the procedure for detecting and tracking atoms using TrackMate in Fig.~S21. To verify our approach we compare the time-dependent diffusion coefficient of our full data-set with the diffusion coefficient obtained from data with a low concentration of atoms in Fig.~S22. Both data sets show similar values for diffusion coefficients and similar time-dependent behavior, showing that this approach is valid. 

It is challenging for TrackMate to identify the moving W atoms on background of a WSe$_2$ lattice, i.e., at the BN/WSe$_2$ interface. For this interface, we tracked the W atoms manually using imageJ. For the pinning site statistics of the BN/WSe$_2$ interface shown in Fig.~\ref{fig:diff-BN-WSe2-v2}(f) we overlaid images such as the one shown in Fig.~\ref{fig:diff-BN-WSe2-v2}(e) with a voronoi lattice which defines the spatial extent of each W, Se$_2$, and H site and we count W atoms that fall within each voronoi cell. 

\subsection{Calculation of Diffusion Coefficients}
Let $\textbf{r}^{(j)}(t) = \{x^{(j)}(t), y^{(j)}(t)\}$
denote the trajectory of Atom~\#\textit{j}.
We track individual atoms using videos obtained with HAADF-STEM imaging with a frame time of $\Delta t$=0.217~s. 
This resulted in time-lapse recorded coordinates on the trajectory, $\textbf{r}^{(j)}_{n} = \textbf{r}^{(j)}(t_n)$ where $t_n=n\Delta t$, $n=0,1,\dots, N$. With this notation, a trajectory is built by $N$ consecutive displacements.

The zero-point in time for the BN/BN and BN/vacuum interface is the start of a new video being recorded in an area that has not previously been imaged. For the BN/WSe$_2$ interface the zero-point in time is the start of each trajectory. We further estimate that approximately 60 s was required to observe moving W atoms when imaging a bulk area, and therefore we plot the time-dependent diffusion coefficient of W atoms at the BN/WSe$_2$ interface starting from t=60 s (Fig.~6(d)). 

We estimate the time-dependent diffusion coefficient 
using CVE, the co-variance based estimator of~\cite{vestergaard2014optimal}, on a rolling time window of 15~s. All  displacements in $x$ and $y$ occurring in a given time interval are used for the estimate. The window is then moved one time-step, $\Delta t$=0,217~s, and the calculation repeated. CVE estimates the diffusion coefficient with~\cite[Eq.~14]{vestergaard2014optimal}
\begin{equation}
    D = \frac{\overline{(\Delta x_n)^2}}{2\Delta t} + \frac{\overline{\Delta x_{n}\Delta x_{n+1}}}{\Delta t} \enspace,
\end{equation}
where $\overline{(\dots)}$ denotes average over all displacements $\Delta x_n$ within a given time window. The error bars in Fig.~\ref{fig:MSD}(d) are given by the square root of the variance of $D$, which is estimated with~\cite[Eq.~17]{vestergaard2014optimal}
\begin{equation}
    \mathrm{var}(D) = D^{2}\left(\frac{6+4\epsilon+2\epsilon^2}{N_{\mathrm{win}}} + \frac{4(1+\epsilon)^2}{N_{\mathrm{win}}^2} \right) \enspace,
\end{equation}
where $N_{\mathrm{win}}$ is the number of displacements in a given time window and $\epsilon = \sigma^2/D\Delta t - 2R$. 
Here, $R $ is the co-called motion blur coefficient, and  $\sigma^2$ is the variance of the localization error. Our images are 512$\times$512 pixels large, and we estimate the size of one W atom to be approximately 20$\times$20 pixels for the magnification we have used. Hence, we estimate $R$ = 20/512 $\approx$ 0.04. The variance $\sigma^2$ is estimated by~\cite[Eq.~15]{vestergaard2014optimal}
\begin{equation}
    \sigma^2 = R\overline{(\Delta x_n)^2} + (2R-1)\overline{\Delta x_n \Delta x_{n+1}}.
\end{equation}

\subsection{DFT Calculations}
Density functional theory (DFT) calculations were performed using the Vienna ab initio Simulation Package (VASP)~\cite{kresse1993ab,kresse1999ultrasoft} within the generalized gradient approximation (GGA) with Perdew–Burke–Ernzerhof (PBE)~\cite{Perdew1996generalized} functional and a plane-wave cut-off of 520 eV and $\Gamma$ point integration in reciprocal space~\cite{kresse1996phys}.
Dispersion correction was described by the Grimme's DFT-D2 method~\cite{grimme2006semiempirical} and we set the global scaling factor $s_6=0.15$ to find the best match of spacing between the BN layers. We also tested the rev-vdW-DF2, optB88, and r2scan functionals and found only a 2\% deviation in the interlayer spacing compared to that found with Grimme's D2 method, see Fig.~S23.
For the BN/WSe$_2$ interface we also used Grimme's DFT-D2 method to find an interlayer distance of 5.1~Å. However, the experimentally measured interlayer distance for a BN/WSe$_2$ heterostructure is about 0.9~Å larger than that predicted by DFT \cite{rooney2017observing}. Therefore, we also tested this larger interlayer spacing, see Fig.~S7(b).

For simplicity, for the BN/WSe$_2$ interface we simulate a BN/WSe$_2$ heterostructure without any rotation between the two crystals. 
Also for simplicity, we only considered B vacancies at the BN/BN interface: based on our other calculations we expect similar results if we add N vacancies. 

We constructed the bilayer BN moiré superlattice with a twisting angle of 21.79$^\circ$ instead of the measured exact 23.9$^\circ$, for the purpose of finding the integer solutions of the moi\'re lattice parameters \cite{feuerbacher2021moire}.
The moiré unit cell consists of 14 B and 14 N atoms, from which we build a $3\times3$ supercell to minimize the interactions between a defect and its images.
For the WSe$_2$/BN bilayer we introduced uniform strain on the BN lattice and constructed a $3\times3$ WSe$_2$ and $4\times4$ BN commensurate supercell.
We use a slab geometry with a 2 nm vacuum layer perpendicular to the atomic plane to minimize the interactions between images. 
All atoms in the supercell were allowed to relax until the residual force per atom was less than 0.01~eV\AA$^{-1}$.

To better estimate how different vacancy type affects the diffusion behavior of W, we calculate the defect formation energy for a W atom on top of either B or N vacancy. The defect formation energy, $E_{\mathrm{f}}$, is given by \cite{wang2021native}

\begin{equation*}
    E_{\mathrm{f}}(W_{\mathrm{B,\,N}}) = E(W_{\mathrm{B,\,N}})- E_{\mathrm{perf}} + \mu_{\mathrm{B,\,N}} - \mu_{\mathrm{W}} + Q(E_{\mathrm{F}}+E_{\mathrm{v}}) ,
\end{equation*}
where $E(W_{\mathrm{B,\,N}})$ is the total energy of a relaxed BN supercell with a W atom and a B or N vacancy, $E_{\mathrm{perf}}$ is the total energy of a perfect BN supercell, $\mu_{\mathrm{B,\,N}}$ is the chemical potential of B or N, $\mu_{\mathrm{W}}$ is the chemical potential of W, $Q$ is the charge state of the vacancy, $E_F$ is the Fermi energy with 0 $<$ $E_F$ $<$ $E_{\mathrm{gap}}$, and $E_{\mathrm{v}}$ is the energy of the top of the valence band. We note that the calculation of defect formation energies is a non-trivial problem in gapped materials, and here for simplicity we did not consider the charged defect, i.e. $Q$=0. The most important variables are the chemical potentials. In our calculation, we choose the elementary limit, meaning that we set $\mu_{\mathrm{B,\,N}}$ equal to the elementary limit for a B or N-rich case, respectively. Therefore, combining $\mu_{\mathrm{B}} + \mu_{\mathrm{N}} =  E_{\mathrm{perf}}$, enables solving the chemical potential $\mu_{\mathrm{B,\,N}}$ in different conditions and thus the defect formation energy.

\section{Contributions}
J.D.T.\ fabricated the samples, performed the STEM experiments and analyzed the data. Y.W.\ performed the DFT simulations. H.F.\ advised on the quantitative diffusion analysis. K.W.\ and T.T.\ grew the BN crystals. E.P.\ acquired overview STEM and TEM images of the samples. P.N.\ and F.M.R.\ supervised the project. J.D.T.\ wrote the manuscript with the input of all co-authors.

\section{Acknowledgements}
This work is partially supported by the U.S. Department of Energy, Office of Science, Basic Energy Sciences (BES), Materials Sciences and Engineering Division under FWP ERKCK47 `Understanding and Controlling Entangled and Correlated Quantum States in Confined Solid-state Systems Created via Atomic Scale Manipulation'.
J.D.T. is partially supported by the Army Research Office MURI (Ab-Initio Solid-State Quantum Materials) grant number W911NF-18-1-0431. P.N. is a Moore Inventor Fellow through Grant GBMF8048 from the Gordon and Betty Moore Foundation. Calculations were performed using resources from the Department of Defense High Performance Computing Modernization program (HPCMP). Additional calculations were performed using resources of the National Energy Research
Scientific Computing Center, a DOE Office of Science User Facility, as well as resources at UCLA.
Transmission electron microscopy was performed using the MIT.nano Characterization Facilities. 
Y.W. acknowledges support from the Chinese Academy of Sciences (Nos.~YSBR047 and E2K5071).
H.F. acknowledges support from the Strategic Programme Excellence Initiative at the Jagiellonian University.
E. P. acknowledges support from a Mathworks Fellowship.
K.W. and T.T. acknowledge support from the JSPS KAKENHI (Grant Numbers 20H00354, 21H05233 and 23H02052) and World Premier International Research Center Initiative (WPI), MEXT, Japan.

\bibliography{references}
\bibliographystyle{naturemag}%

\end{document}